\documentclass[12pt]{article}
\begin{document}
\begin{center}
{\large { Lie Particle And Its Batalin-Tyutin Extension }} \vskip
2cm
Subir Ghosh\\
\vskip .3cm
Physics and Applied Mathematics Unit,\\
Indian Statistical Institute,\\
203 B. T. Road, Calcutta 700108, India\\
E-mail address: sghosh@isical.ac.in \\
\vskip .3cm
And\\
\vskip .3cm
Probir Pal\\
Physics Department,\\
Uluberia College,\\
Uluberia, Howrah 711315,  India.\\
E-mail address: probirkumarpal@rediffmail.com \\
\end{center}

\vskip 3cm
 {\bf{Abstract:}}\\
 In this Letter we have proposed a point particle model that generates a
 noncommutative three-space, with the coordinate brackets being Lie algebraic
 in nature, in particular isomorphic to the angular momentum algebra. The work is in the
 spirit of our earlier works in this connection,
 {\it {i.e.}} PLB 618 (2005)243 and PLB 623 (2005)251, where the $\kappa $-Minkowski form
 of noncomutative spacetime was considered. This non-linear and
 operatorial nature of the configuration space coordinate algebra can pose problems
 regarding its quantization. This prompts us to embed the model in the Batalin-Tyutin
 extended space where the equivalent model comprises of phase space variables satisfying
 a canonical algebra. We also compare our present model with the  point particle model,
 previously proposed by us, in the context of $\kappa$-Minkowski spacetime.

\newpage
I. {\it {Introduction}}: \vskip .5cm \noindent

In the present Letter, we continue to study general quantum spaces
along the lines proposed in \cite{s1,s2}. In particular, we
provide a Hamiltonian system, in the point particle approach
\cite{s1,s2} that simulates a Non-Commutative (NC) space where the
operatorial form of NC has a Lie algebraic structure,
\begin{equation}
\{x^i,x^j\}=\kappa \epsilon^{ijk}x^k~,~~\{x^i,t\}=0,
\label{lie}\end{equation} where $\epsilon^{ijk}$ is the fully
anti-symmetric Levi-Civita tensor, with $\epsilon^{123}=1$. Here
$x^i$ denotes the three dimensional configuration space and
$\kappa$ is the NC parameter. Notice that Jacobi identity
corresponding to this NC algebra is satisfied. This, in fact, is a
weaker condition than the criteria of noncommutativity set by
\cite{dfr}, where double commutators in the configuration space
algebra are supposed to vanish.

Indeed, NC spaces (or more generally NC spacetimes) are   here to
stay (for reviews see \cite{rev}). It all started in  quantum
field theory when Snyder \cite{sny} introduced NC spacetime to
regularize the short distance singularity in a Lorentz covariant
way. The recent excitement in NC spacetime physics is generated
from the seminal work of Seiberg and Witten \cite{sw} who
explicitly demonstrated  the emergence of NC manifold in certain
low energy limit of open strings moving in the background of a two
form gauge field. In the quantum gravity scenario, very general
considerations in black hole physics lead to the notion of a fuzzy
or Non-Commutative (NC) spacetime which can avoid the paradoxes
one faces in trying to localize a spacetime point within the
Planck length \cite{rev}. This is also corroborated in the
modified Heisenberg uncertainty principle that is obtained in
string scattering results. In these instances, the NC spacetime is
expressed by the Poisson bracket algebra (to be interpreted as
commutators in the quantum analogue),
\begin{equation}
\{x^\mu,x^\nu\}=\theta ^{\mu\nu}, \label{nc0} \end{equation} where
noncommutativity is of the simplest form, $\theta ^{\mu\nu}$ being
a $c$-number constant. From the above well known examples, it
might appear that NC physics is relevant only in relativistic
quantum theories and that too at the Planck regime. However, this
is far from the truth as the recent shift of focus of NC physics
in condensed matter systems clearly indicates. NC Chern Simons
theory as an effective field theory for fractional quantum effect
has been established long ago \cite{suss}. The latest excitement
is in the context of an observable evidence  of a magnetic
monopole structure in crystal momentum space of Bloch electrons,
that has appeared in the works of Onoda and Nagaosa \cite{mono1}
and Fang et al. \cite{mono2}.  These results have found an elegant
interpretation in the  NC phase space models Berard and Mohrbach
\cite{berard}. A recent work by Duval et al. \cite{duval} have
explained Berry curvature effect in Bloch electron dynamics
suggested by Xiao et al. \cite{xiao} through an NC phase space.
The relevance of these ideas in the context of {\it{spintronics}}
\cite{spin}, a spin Hall effect which might induce a spin current
in semiconductor samples, has been elaborated in \cite{shin}.

It is very significant that the NC effects in condensed matter
systems have brought in to fore their relevance in
non-relativistic physics and at the same time have emphasized the
importance of more general ({\it{operatorial}}) forms of
noncommutativity that is present in the latter systems. This is
quite in contrast to the simple constant form of NC spacetime
(\ref{nc0}) that was prevalent before.

 However, the theoretical framework for    NC spacetimes, with a  Lie algebraic
 form of structure constants $C^{\mu\nu}_\lambda$
 \begin{equation}
\{x^\mu,x^\nu\}=C^{\mu\nu}_\lambda x^\lambda  \label{nc1}
\end{equation}
has already been developed in \cite{maj}. It is important to note
that the  NC extension in \cite{dfr,g1} is {\it{operatorial}} and
do not jeopardize the Lorentz invariance in relativistic models,
which is the case with (\ref{nc0}) with constant $\theta
^{\mu\nu}$. (For an introduction to this subject the readers are
referred to \cite{ga}.) Of particular importance in the above is a
restricted class of spacetimes known as $\kappa $-Minkowski
spacetime (or $\kappa $-spacetime in short), that is described by
the algebra,
\begin{equation}
\{x^i,t\}=kx^i ~~,~~ \{x^i,x^j\}=\{t,t\}=0 .\label{kmin}
\end{equation}
In the above, $x^i$ and $t$ denote the space and time operators
respectively. Some of the important works in this topic that
discusses, among other things, construction of a quantum field
theory in $\kappa $-spacetime, are provided in \cite{g1}. Very
interestingly, Amelino-Camelia \cite{gam} has proposed an
alternative path to quantum gravity - "the doubly special
relativity" - in which {\it two} observer independent parameters,
(the velocity of light and Planck's constant), are present. It has
been shown \cite{kowal} that $\kappa$-spacetime is a realization
of the above. Our previous works \cite{s1,s2} along these lines
provide a point particle model for the above spacetime. Indeed, it
has been rightly pointed out by Amelino-Camelia in \cite{ga} that
of all the different forms of Lie algebraic NC structures,
$\kappa$-Minkowski algebra (\ref{kmin}) is probably the simplest.
This fact will also be amply demonstrated at the end of our work,
where we compare and contrast between the present particle model
and the one formulated for $\kappa$-spacetime in \cite{s1,s2}.

Let us put our work in the proper perspective. Since the
operatorial forms NC spaces (or times) are playing  increasingly
dominant roles in non-relativistic (condensed matter and other)
systems which are accessible to experimental verifications, it
becomes imperative that one should have an intuitive (particle
like) model that is equivalent to the above in some sense. In a
non-relativistic setup, NC space, originating from the lowest
Landau level projection of charged particles moving in a plane
under the influence of a uniform, perpendicular (and strong)
magnetic field \cite{rev}, has become the prototype of a simple
physical system (qualitatively) describing considerably more
complex and abstract phenomena, in this case open strings moving
in the presence of a background two form gauge field \cite{sw}
mentioned before. This sort of intuitive picture, if present, is
very useful and appealing. We provided this mechanism in
\cite{s1,s2} for the $\kappa$-Minkowski spacetime. The present
work deals with the Lie algebraic form of NC space as given in
(\ref{lie}).

The model we propose has Hamiltonian constraints in the Dirac
formalism \cite{dir} (or equivalently it has a non-trivial
symplectic structure \cite{fad}). The resulting Dirac brackets
\cite{dir} provide the NC space algebra in a classical setup.
Quite obviously, the non-linear and operatorial form of Dirac
brackets obtained here are not amenable to a conventional
quantization where one elevates the status of Dirac brackets to
commutators via the correspondence principle. This problem is
tackled in the Batalin-Tyutin (BT) formalism \cite{bt} where one
embeds the physical system in a suitably extended phase space such
that one has a dual system in a completely {\it{canonical}} phase
space. This removes the ambiguities in the prescription for
canonical quantization in this type of models. The BT embedding is
quite involved due to the non-linear nature of the constraints and
we explicitly discuss the results for $\kappa$ and BT auxiliary
Degrees Of Freedom (DOF) in the lowest  non-trivial order. Other
examples of BT embedding for non-linear systems can be found in
\cite{sg}.

The paper is organized as follows: In Section II  we  put forward
the non-relativistic particle model that has an underlying
symplectic structure that simulates the NC coordinate space
algebra, isomorphic to the angular momentum Lie algebra. Section
III is devoted to the BFT embedding of our model.  In Section IV
we  put forward a comparison between the proposed particle models
(in this paper as well as in the previous ones \cite{s1,s2})
referring to different Lie algebraic structures of
noncommutativity. The paper ends with a Conclusion in Section V.

\vskip 1cm {\it II. {The Lie particle}}: \vskip .5cm \noindent We
posit the following first order Lagrangian
\begin{equation}
L=\dot\eta^k(x^k+\frac{\kappa}{2}\epsilon^{ijk}x^i\eta^j)-V,
\label{01}
\end{equation}
describing the "Lie particle" that generates the spatial angular
momentum algebra we are interested in. The potential $V$ is so far
unspecified. The phase space variables are
\begin{equation}
p^i=0~;~~\pi^i=x^i+\frac{\kappa}{2}\epsilon^{ijk}x^j\eta^k,
 \label{03}
\end{equation}
with the Poisson brackets
\begin{equation}
\{x^i,p^j\}=
\{\eta^i,\pi^j\}=\delta^{ij}~,~~\{(x^i,p^i),(\eta^j,\pi^j)\}=0.
\label{02}
\end{equation}

Let us quickly review the Hamiltonian constraint analysis as
formulated by Dirac \cite{dir}. In this scheme the constraints are
termed as First Class Constraints (FCC) if they commute modulo
constraints (in the Poisson Bracket  sense)
  or Second Class Constraints (SCC)
if they do not. The FCCs induce gauge invariance in the theory
whereas the SCCs tend to modify the symplectic structure of the
phase space for compatibility with the SCCs. The above
modification induces a replacement of the Poisson brackets by
Dirac Brackets (DB) as defined below,
\begin{equation}
\{A,B\}_{DB}= \{A,B\}-\{A,\psi _\alpha \} (\Psi
_{\alpha\beta})^{-1}\{\psi _\beta ,B\}. \label{dirac}
\end{equation}
where $\psi _\alpha $ refer to the SCCs and $\Psi _{\alpha \beta
}\equiv \{\psi _{\alpha },\psi _{\beta }\}$ is invertible on the
constraint surface. The Dirac brackets are compatible with the
SCCs so that the SCCs can be put "strongly" to zero.

According to the above classification \cite{dir} the set of
constraints obtained from (\ref{03}),
\begin{equation}
\psi^{i}_{1}=p^{i}~~,~~\psi^{i}_{2}=\pi^{i}-x^{i}-\frac{\kappa}{2}\epsilon^{ijk}x^{j}\eta^{k},
\label{1}
\end{equation}
comprise  an SCC system  since they obey a non-vanishing Poisson
bracket structure given by,
\begin{equation}
\Psi _{\alpha \beta }^{ij}=\{\psi _{\alpha }^{i},\psi _{\beta
}^{j}\}~;~~\alpha ,\beta \equiv 1,2 \label{11a}
\end{equation}
where
$$
\Psi _{\alpha \beta }^{ij}=\left(
\begin{array}{cc}
0 &   \delta^{ij}-\frac{\kappa}{2}\epsilon^{ijk}\eta^{k}\\
-\delta^{ij}-\frac{\kappa}{2}\epsilon^{ijk}\eta^{k} &
\kappa\epsilon^{ijk}x^{k}
\end{array}
\right).
$$
The inverse is computed in a straightforward way,
$$
\Psi _{\alpha \beta }^{(-1)ij}=\left(
\begin{array}{cc}
\frac{\kappa}{1+\frac{\kappa^{2}}{4}\eta^{2}}(\epsilon^{ijk}x^{k}+\frac{\kappa}{2} (\eta^{i}x^{j}-\eta^{j}x^{i})) &   -\frac{1}{1+\frac{\kappa^{2}}{4}\eta^{2}}(\delta^{ij}-\frac{\kappa}{2}\epsilon^{ijk}\eta^{k}+\frac{\kappa^{2}}{4}\eta^{i}\eta^{j})\\
\frac{1}{1+\frac{\kappa^{2}}{4}\eta^{2}}(\delta^{ij}-\frac{\kappa}{2}\epsilon^{ijk}\eta^{k}+\frac{\kappa^{2}}{4}\eta^{i}\eta^{j})
& 0
\end{array}
\right).
$$
From the definition (\ref{dirac}) this leads to the Dirac
brackets,
\begin{equation}
\{x^{i},x^{j}\}_{DB}=\frac{\kappa}{1+\frac{\kappa^{2}}{4}\eta^{2}}(\epsilon^{ijk}x^{k}+\frac{\kappa}{2}
(\eta^{i}x^{j}-\eta^{j}x^{i})), \label{2}
\end{equation}
$$
\{\eta^{i},\pi^{j}\}_{DB}=\delta^{ij}~;~~\{\eta^{i},\eta^{j}\}_{DB}=0,
$$
$$
\{x^{i},\eta^{j}\}_{DB}=-\frac{1}{1+\frac{\kappa^{2}}{4}\eta^{2}}(\delta^{ij}-\frac{\kappa}{2}\epsilon^{ijk}\eta^{k}+\frac{\kappa^{2}}{4}\eta^{i}\eta^{j})
$$
\begin{equation}
\{x^{i},\pi^{j}\}_{DB}=\frac{\kappa}{2(1+\frac{\kappa^{2}}{4}\eta^{2})}(\epsilon^{ijk}x^{k}+\frac{\kappa}{2}
((x.\eta )\delta^{ij}-x^{i}\eta^{j})-\frac{\kappa
^2}{4}\eta^{i}\epsilon^{jkl}\eta^{k}x^{l}). \label{3}
\end{equation}
In order to focus on the salient features we restrict ourselves to
results valid up to  the lowest non-trivial order in $\kappa$ and
obtain the NC algebra
\begin{equation}
\{x^i,x^j\}_{DB}=\kappa\epsilon^{ijk}x^k+O(\kappa ^2)~,
\label{005}
\end{equation}
$$
\{x^i,\eta^j,\}_{DB}=-\delta^{ij}+\frac{\kappa}{2}\epsilon^{ijk}\eta^k+O(\kappa
^2)$$
\begin{equation}
\{x^i,\pi^j \}_{DB}=\frac{\kappa}{2}\epsilon^{ijk}x^k+  O(\kappa
^2);~ \{\eta^i,\eta^j\}_{DB}=0
~;~\{\eta^i,\pi^j\}_{DB}=\delta^{ij}+O(\kappa ^2) \label{05}
\end{equation}
Notice that to $O(\kappa)$ we have recovered the angular momentum
algebra for the coordinates. This is one of our main results.

Let us note an important point related to the applicability of our
formalism for the general Lie algebraic structure, as given in
(\ref{nc1}). It is quite clear that the presence of
$\epsilon^{ijk}$ in the  $\kappa $-term of the Lagrangian in
(\ref{01}) is responsible for the cherished Dirac Bracket of
(\ref{005}):
$$
\{x^i,x^j\}_{DB}=\kappa\epsilon^{ijk}x^k+O(\kappa ^2).
$$
In order to reproduce the general form of Lie algebra among $x^i$,
as given in (\ref{nc1}), one has to modify the Lagrangian
accordingly. Hence, in principle, there is no problem in
generalizing our model for the general Lie algebraic
noncommutativity.

We fix the form of potential $V$ to get an idea about the dynamics
and a natural choice is to consider a harmonic oscillator
potential,
\begin{equation}
H=V=x^2+\nu\eta^2
 \label{001}
\end{equation}
which is scaled by an overall  dimensional parameter apart from
the constant $\nu$. The equations of motion for a generic variable
$O$ is
\begin{equation}
\dot O=\{O,H\}_{DB} .\label{06} \end{equation} In the present case
we find
\begin{equation}
\dot x^i=-2\nu\eta^i~,~~\dot
\eta^i=2x^i-\kappa\epsilon^{ijk}x^j\eta^k .\label{07}
\end{equation}
Iterating the above equations one more time we get
\begin{equation}
\ddot x^i=-4\nu x^i+2\nu\kappa\epsilon^{ijk}x^j\eta^k
~,~~\ddot\eta^i=-4\nu\eta^i . \label{08}
\end{equation}
Note that we will get harmonic oscillator equations of motion for
both $x^i$ and $\eta^i$ only if $\nu =\kappa $.

It is well-known from the theory of differential forms (Darboux's
theorem) that for a symplectic manifold, by a suitable
transformation locally one can always go to a set of coordinates
which are canonical. In the present case associativity, that is
validity of the Jacobi identity, among phase space variables is
assured since we have used Dirac brackets. The Darboux coordinates
are defined to be,
\begin{equation}
Q^i\equiv x^i+\frac{\kappa}{2}\epsilon^{ijk}x^j\eta
^k~;~~P^i\equiv -\eta ^i . \label{dar}
\end{equation}
They obey the canonical algebra
\begin{equation}
\{Q^i,Q^j\}=\{P^i,P^j\}=0~;~~\{Q^i,P^j\}=\delta^{ij}. \label{dar1}
\end{equation}

\vskip 1cm III. {\it  {BT embedding of the Lie particle}}: \vskip
.5cm \noindent We start by a brief  digression  on the BT
formalism \cite{bt}. The basic idea is to embed the original
system in an enlarged phase space (the BT space), consisting of
the original "physical" degrees of freedom and auxiliary
variables, in a particular way such that the resulting enlarged
system possesses local gauge invariance. Imposition of gauge
conditions accounts for the true number of degrees of freedom and
at the same time the freedom of having different gauge choices
leads to structurally distinct systems. However, all of them are
assured to be gauge equivalent.

Let us consider a generic set of constraints $(\psi _\alpha ,\xi
_l )$ and a Hamiltonian operator $H$ with the following PB
relations,
$$
\{\psi _\alpha  (q) ,\psi _\beta (q)\}\approx \Delta _{\alpha
\beta
 }(q) \ne 0~~;~~\{\psi _\alpha (q) ,\xi _l (q)\}\approx 0
$$
\begin{equation}
 \{\xi _l(q) ,\xi _n (q)\}\approx 0 ~~;~~
\{\xi _l(q) ,H (q)\}\approx 0. \label{011}
\end{equation}
In the above $(q)$ collectively refers to the set of variables
present prior to the BT extension and "$\approx $" means that the
equality holds on the constraint surface. Clearly $\psi _\alpha  $
and $\xi _l $ are SCC and FCC  respectively.

In systems with non-linear SCCs, (such as the present one), in
general the DBs can become dynamical variable dependent due to the
$\{A,\psi _\alpha \}$ and $\Delta _{\alpha \beta }$
 terms, leading to problems for
the quantization programme. This type of pathology is cured in the
BT formalism in a systematic way, where one introduces the BT
variables $\phi
 ^\alpha  $, obeying
\begin{equation}
\{\phi ^\alpha ,\phi ^\beta \}=\omega ^{\alpha \beta}= -\omega
^{\beta \alpha}, \label{bt}
\end{equation}
where $\omega ^{\alpha \beta}$ is a constant (or at most
 a c-number function) matrix, with the aim of modifying the SCC
$\psi _\alpha  (q)$ to $\tilde \psi _\alpha  (q, \phi ^\alpha )$
such that,
\begin{equation}
\{\tilde\psi _\alpha (q,\phi ) ,\tilde\psi _\beta (q,\phi )\}=0
~~;~~\tilde\psi _\alpha (q,\phi )=\psi _\alpha (q)+ \Sigma
_{n=1}^\infty \tilde\psi ^{(n)} _\alpha (q,\phi )~~;~~ \tilde\psi
^{(n)}\approx O(\phi ^n) \label{b1}
\end{equation}
This means that $\tilde\psi  _\alpha $ are now FCCs and in
particular abelian. The explicit terms in the above expansion are,
\begin{equation}
\tilde\psi ^{(1)}_\alpha =X_{\alpha \beta }\phi ^\beta ~~;~~
\Delta _{\alpha \beta }+X_{\alpha \gamma } \omega ^{\gamma \delta
}X_{\beta \delta }=0 \label{b2}
\end{equation}

\begin{equation}
\tilde\psi ^{(n+1)}_\alpha =-{1\over{n+2}} \phi^{\delta }\omega
_{\delta \gamma }X^{\gamma \beta }B^{(n)}_{\beta \alpha }~~;~~n\ge
1 \label{b3}
\end{equation}

\begin{equation}
B^{(1)}_{\beta \alpha }= \{\tilde\psi ^{(0)} _\beta ,\tilde\psi
^{(1)} _\alpha \}_{(q)}- \{\tilde\psi ^{(0)} _\alpha ,\tilde\psi
^{u(1)} _\beta \}_{(q)} \label{b4}
\end{equation}

\begin{equation}
B^{(n)}_{\beta \alpha }= \Sigma _{m=0}^n \{\tilde\psi ^{(n-m)}
_\beta ,\tilde\psi ^{(m)} _\alpha \}_{(q,p)}+ \Sigma _{m=0}^n
\{\tilde\psi ^{(n-m)} _\beta ,\tilde\psi ^{(m+2)} _\alpha
\}_{(\phi )} ~~;~~n\ge 2 \label{b5}
\end{equation}
In the above, we have defined,
\begin{equation}
X_{\alpha \beta }X^{\beta \gamma }= \omega _{\alpha \beta }\omega
^{\beta \gamma } =\delta ^\gamma _\alpha   . \label{b6}
\end{equation}
A very useful idea is to introduce the improved function $\tilde
f(q)$
 corresponding to each $f(q)$,
\begin{equation}
\tilde f(q,\phi )\equiv f(\tilde q) =f(q)+\Sigma _{n=1}^\infty
\tilde f(q,\phi )^{(n)}~~ ;~~\tilde f^{(1)}=- \phi^{\beta }\omega
_{\beta \gamma }X^{\gamma \delta }\{ \psi_\delta ,f\}_{(q)}
\label{b7}
\end{equation}

\begin{equation}
\tilde f^{(n+1)}=-{1\over{n+1}} \phi^{\beta }\omega _{\beta \gamma
}X ^{\gamma \delta } G(f)^{\lambda (n)}_\delta ~~;~~n\ge 1
\label{b8}
\end{equation}

\begin{equation}
G(f)^{(n)}_{\beta }= \Sigma _{m=0}^n \{\tilde\psi ^{ (n-m)} _\beta
,\tilde f^{(m)}\}_{(q)}+ \Sigma _{m=0}^{(n-2)} \{\tilde\psi ^{
(n-m)} _\beta ,\tilde f^{(m+2)}\}_{(\phi )} +\{\tilde\psi ^{
(n+1)} _\beta ,\tilde f^{(1)}\}_{(\phi )} \label{b9}
\end{equation}
which have the property $\{\tilde\psi _\alpha (q,\phi ) ,\tilde
f(q,\phi )\}=0$. Thus, in the BT space, the improved functions are
FC or equivalently gauge invariant. Note that $\tilde q$
corresponds to the improved variables for $q$. The subscript
$(\phi )$ and $(q)$ in the PBs indicate the  variables with
respect to which the PBs are to be taken. It can be proved that
extensions of the original FCC $\xi _l $ and Hamiltonian
 $H$ are simply,
\begin{equation}
\tilde \xi _l=\xi (\tilde q)~~;~~ \tilde H=H (\tilde q).
\label{b10}
\end{equation}
One can also reexpress the converted SCCs as $\tilde\psi
^\mu_\alpha \equiv \psi ^\mu_\alpha (\tilde q)$. The following
identification theorem holds,
\begin{equation}
\{\tilde A,\tilde B \}=\tilde {\{A,B\}_{DB}}~~;~~ \{\tilde
A,\tilde B \}\mid _{\phi =0}=\{A,B \}
 _{DB}~~;~~\tilde 0=0.
\label{b11}
\end{equation}
Hence the outcome of the BT extension is the closed system of FCCs
with the FC Hamiltonian given below,
\begin{equation}
\{\tilde \psi _\alpha  ,\tilde \psi _\beta \}= \{\tilde \psi
_\alpha  ,\tilde \xi _l\}= \{\tilde \psi _\alpha ^\mu ,\tilde H\}=
0~~;~~ \{\tilde \xi _l ,\tilde \xi _n\}\approx 0 ~;~ \{\tilde \xi
_l ,\tilde H\}\approx 0. \label{b12}
\end{equation}
In general, due to the non-linearity in the SCCs, the extensions
 in the improved variables, (and subsequently in the FCCs
 and FC Hamiltonian), may
turn out to be infinite series. This type of situation has been
encountered
 before \cite{sg}.

 Let us concentrate on the problem at hand. The BT variables satisfy the canonical algebra
\begin{equation}
\{\phi_{i\alpha },\phi_{j\beta
}\}=\delta_{ij}\epsilon_{\alpha\beta }\equiv \omega_{ij\alpha\beta
}~;~~\epsilon^{01}=-\epsilon_{01}=1. \label{b22}
\end{equation}
The structure of $\tilde O$ for a generic variable $O$ is of the
form
\begin{equation}
\tilde O=O+O^{(1)}(\phi)+O(\phi\phi ,...). \label{bb3}
\end{equation}
Truncating the above up to a single BT variable contribution we
compute the FC counterparts of all the degrees of freedom:
$$\tilde x^i=x^i-\phi^{1i}+\frac{\kappa }{4}\epsilon^{ijk}\phi ^{1j}\eta
^k+\frac{\kappa}{2}\epsilon^{ijk}\phi^{2j}x^k$$
$$\tilde \eta^i=\eta^i+\phi^{2i}+\frac{\kappa }{4}\epsilon^{ijk}\phi ^{2j}\eta
^k $$
\begin{equation}
\tilde p^i=p^i+\phi^{2i}-\frac{\kappa }{4}\epsilon^{ijk}\phi
^{2j}\eta ^k ~,~~\tilde
\pi^i=\pi^i+\frac{\kappa}{2}\epsilon^{ijk}\phi^{2j}x^k \label{b33}
\end{equation}
The corresponding FC Hamiltonian will be
\begin{equation}
\tilde H=\tilde x^2+\nu\tilde \eta ^2. \label{b34}\end{equation}
This Hamiltonian together with the canonical set of phase space
variables in the BT extended space can be quantized in the
conventional way.

\vskip 1cm IV. {\it  {Comparison between Lie particle models}}:
\vskip .5cm \noindent In this section, we will now compare and
contrast some of the features of the point particle model in
\cite{s1} for the $\kappa $-spacetime and the particle model
formulated here. In both cases, we started with an extended phase
space having a single scalar variable $\eta $ in \cite{s1} and a
vector $\eta ^i$ in the present model, as auxiliary degrees of
freedom. However, in \cite{s1}, in the reduced space, one can
incorporate the constraints strongly and completely eliminate
$\eta $ with the resulting system comprising of only
$(X^i,P^j)$-variables. In the present model, there is no simple
way of removing $\eta ^i$ due to the more involved constraint
structure, as derived in (\ref{1}).

Secondly, in the $\kappa $-Minkowski algebra (\ref{kmin}), a
possible set of Darboux transformation can be,
\begin{equation}
x^i\equiv X^i~,~~t\equiv \kappa (X^iP^i)~,~~  p^i\equiv P^i ,
\label{c1}
\end{equation}
with $X^i$ and $P^j$ obeying canonical brackets. One can check
that all the Jacobi identities are preserved {\footnote{The only
non-trivial one being
$J(x^i,p^j,t)=\{\{x^i,t\},p^j\}+\{\{t,p^j\},x^i\}+\{\{p^j,x^i\},t\}=0$.}}.
This shows that $X^i$ and $P^j$ constitute a consistent set of
degrees of freedom. Incidentally, the angular momentum operator
also remains unchanged.

But these features are not preserved for the Lie algebra
(\ref{lie}). In fact Jacobi identities in the $(X^i,P^j)$-phase
space get violated. This reestablishes the fact that the extension
by $\eta^i $ is necessary and one can not simply remove $\eta ^i$
and get a consistent set of variables with a symplectic algebra.
For the same reason, the Darboux coordinates contain $\eta ^i$ in
(\ref{dar}) in an entangled way. Also, the angular momentum
operator can be formulated in terms of the Darboux variables but
it will appear quite complicated in terms of original variables.
The other option is to exploit the Batalin-Tyutin framework, (that
we have studied in Section III), where the extended phase space is
canonical and one can construct angular momentum operator (and
other relevant operators) in a straightforward manner.

The above discussion reconfirms the fact that $\kappa $-Minkowski
spacetime is one of the simplest forms of Lie algebraic NC
spacetime and this property is inherited by the corresponding
particle model in \cite{s1}. On the other hand, some of these
pleasant features are lost in the present point particle model
that generates the configuration space algebra (\ref{lie}). \vskip
1cm V. {\it  {Conclusion}}: \vskip .5cm \noindent First of all,
let us summarize our work. We have provided a non-relativistic
point particle model that induces a noncommutative three-space
endowed with Lie algebraic form of commutator brackets. To be
specific, the coordinates obey an angular momentum algebra.  This
type of configuration space that has a non-trivial algebra between
the coordinate variables is not suitable for a canonical
quantization. To circumvent this problem we have considered
instead an extension of the model in Batalin-Tyutin phase space
where a dual model to our original one can be considered in a
canonical phase space. We have also shown how the auxiliary
variables play an essential role here in comparison to other
particle model in \cite{s1} that induce $\kappa $-Minkowski
spacetime.

In recent years, lot of attention is being paid to spaces (or
spacetimes) having a general (operatorial) form of
noncommutativity. Indeed, interesting and exciting systems in the
domain of low energy condensed matter bear interpretations in
terms of effective models that has underlying noncommutative space
structures. We believe that our point particle approach will be
able to provide some intuitive understanding of these models in a
simpler setting. \vskip 1cm  {\it  {Acknowledgement}}: \vskip
.1cm \noindent It is a pleasure to thank the Referee for the
instructive comments. 
\end{document}